\documentstyle[a4,12pt,iopl12]{ioplppt}

\author{}
\article{The Chazy equation}{Instability of pole singularities for the
Chazy equation}
\author{Satyanad Kichenassamy}
\address{
        School of Mathematics,
        University of Minnesota,
        127 Vincent Hall,
        206 Church Street S.\ E.,
        \mbox{} Minneapolis, MN 55455-0487.
Current address:
Max-Planck-Institut f\"ur Mathematik in den Natur\-wissenschaften,
Inselstra\ss e 22-26, D-04103 Leipzig, Germany.
        E-mail: {\tt kichenas\char'100 mis.mpg.de}.}
\date{}
\jl{1}

\abstract
We prove that the negative resonances of the Chazy equation (in the
sense of Painlev\'e analysis) can be related directly to its
group-invariance properties. These resonances indicate in this case
the instability of pole singularities. Depending on the value of a
parameter in the equation, an unstable isolated pole may turn into the
familiar natural boundary, or split into several isolated
singularities. In the first case, a convergent series representation
involving exponentially small corrections can be given. This
reconciles several earlier approaches to the interpretation of
negative resonances.  On the other hand, we also prove that pole
singularities with the maximum number of positive resonances are
stable. The proofs rely on general properties of nonlinear Fuchsian
equations.
\endabstract

\pacs{02.30.Hq}

\newtheorem{th}{Theorem}

\newtheorem{lemma}[th]{Lemma}

\newcommand{\beq}{\begin{equation}}
\newcommand{\eeq}{\end{equation}}
\newcommand{\bea}{\begin{eqnarray*}}
\newcommand{\eea}{\end{eqnarray*}}
\newcommand{\bt}{\begin{th}}
\newcommand{\et}{\end{th}}

\newcommand{\pa}{\partial}

\newcommand{\ep}{\varepsilon}

\renewcommand{\today}{}

\renewcommand{\Box}{
\raisebox{-1pt}{\hbox{\vrule\vbox to 9pt{\hrule width 9pt\vfill\hrule}\vrule\,}} }

\begin{document}
\maketitle



\section{Introduction}

The method of pole expansions is a versatile tool for the
investigation of singularities of ODE and PDE. Going as far back as
the work of Briot and Bouquet, and Kowalewska, this method acquired
new significance when it was found that symmetry reductions of
equations integrable by inverse scattering usually have the Painlev\'e
property in the strong sense: all of their movable singularities are
described by the method of pole expansions \cite{a-s,ars}. Conversely
the inverse scattering transform suggested new approaches to the study
of the Painlev\'e equations themselves.

We are interested in the case of the Chazy equation
\begin{equation}\label{chazy}
y''' - 2yy'' +3y^{\prime 2} = 0,
\end{equation}
where this method apparently fails to provide information on the
general solution of the equation under consideration, because the pole
expansion has fewer free parameters than the order of the equation,
and therefore does not seem to represent the general solution. The
simplest issue is this: the Chazy equation has order three and
possesses the exact solution
\[
-6/x.
\]
However, this is, up to translations, the only solution with a simple
pole. The `pole expansion' reduces in this case to its first term.
This solution should be embedded in a three-parameter family of
solutions: the fact that there is only a one-parameter family of
solutions with simple poles shows that simple poles are {\em unstable
  under perturbation,} but does not give further information. (More
background information on the Chazy equation is given in section 1.3
below.)  Kruskal (see \cite{k-j-h}) suggested that the failure of the
pole expansion was due to the omission of exponentially small terms in
the expansion of the general solution $y(x)$. Furthermore, this
representation should not be valid in a full neighborhood of $x=0$ in
general, but only in a sector, as in the case of expansions near an
irregular singular point. It turns out in this case that the expansion
in powers of exponentials is convergent, which is not expected at an
irregular singularity. On the other hand, the perturbative approach
\cite{f-p,c-f-p}, shows that there is a formal series solution
$y(x;\ep) = y(x) + \ep y_1(x) + \dots$, which does contain the full
number of arbitrary parameters. Also, the perturbative approach
accounts for the apparent paradox that the solutions of the linearized
equation have solutions which are more singular than any actual
solution of the equation: they are related to the variation of free
parameters in the general solution, at a fixed location away from the
singularity. However, all of the terms in the expansion in powers of
$\ep$ have only power singularities, and do not contain exponential
terms. There is also no indication that the actual solution cannot be
continued around the singularity.

Another difficulty comes from the general `class XII' equation of
Chazy:
\begin{equation}\label{xii}
y''' - 2yy'' +3y^{\prime 2} = E (6y'-y^2)^2,
\end{equation}
where $E=4/(36-k^2)$, with $k\geq 0$, omitting the `complementary
terms.'  This equation has the same special solution, but the general
solution can be completely different, and is in fact rational for some
values of $k$. It is possible for the simple pole to split into two or
more poles by perturbation. This effect cannot be captured adequately
in a series representation in powers of $x$: take a solution with two
poles at $x=0$ and $x=\alpha$. The radius of convergence of a pole
expansion around $x=0$ cannot be greater than $|\alpha|$, and
therefore tends to zero if there is a confluence of the two
singularities ($\alpha\to 0$). The perturbative approach can
correctly describe what happens at a fixed location away from the
singularities, but its asymptotics as $x\to 0$ do not describe the
singularity correctly. This can be seen in the treatment of the first
example of section 5 in \cite{c-f-p} where the confluence can only be
ascertained by explicitly summing the pole expansion.

We address these difficulties as follows:
\begin{enumerate}
\item In considering the perturbative solution $y(x;\ep)$, it is not
  permissible to exchange the limits $x\to 0$ and $\ep\to 0$. This
  explains why no exponential terms are generated there.
\item To recover the results of the perturbative approach from the
  exponential expansion, it is necessary to vary the free parameters
  in a very particular way, which we relate to the group invariance of
  the Chazy equation.
\item It is possible to describe confluence phenomena analytically by
  means of a Cole-Hopf transformation: if $u(x;\ep)$ is an analytic
  family of functions where a pair of simple zeros coalesces at $x=0$,
  the family $u'/u$ has a confluence of poles. Note that we can allow
  $u$ to be analytic in both arguments, whereas the pole expansion of
  $u'/u$ would have a vanishingly small radius of convergence, as
  explained above.
\end{enumerate}

It is convenient to describe all of the above issues as relating to
the stability of singular behavior, because of the close similarity to
the stability of solitary waves in nonlinear wave equations. Note that
some authors (in particular Bureau) define a `stable equation' as one
which possesses the Painlev\'e property. We are interested in a
different issue, namely whether the leading asymptotics of a
particular solution of an ODE or PDE are stable under perturbation of
the solution.
Consider for instance real-valued solutions of the equation
\[
u_{tt} - u_{xx} - u_{yy} = e^{u},
\]
which has the special solution $e^u=2/t^2$; then \cite{inversibilite}
all nearby solutions are such that $e^u$ has an expansion in powers of
$T$ and $T\ln T$, where $T=t-\psi(x,y)$, with $\psi$ small, and the
first term is $\ln (2/T^2)$. It is not possible to exclude logarithms
in general, but the leading order of the singularity remains the same.
The singularity of the solution $\ln (2/t^2)$ would be termed `stable'
in the sense of the present paper.

The main technical tool in the proofs of our results is the Fuchsian
algorithm (\cite{syd,gks} and earlier references therein). The basic
step is to show that the method of pole expansions for ODE or PDE
amounts to seeking solutions of a nonlinear ordinary or partial
differential equation with a regular singularity, which may include
logarithmic terms. As far as the present discussion is concerned, note
that several authors (in particular, Bureau, see also Conte, Fordy and
Pickering \cite{c-f-p}) have identified ARS-WTC resonances with the
indices of a linear Fuchsian ODE, and have observed in many cases that
the linearization of the equation itself is Fuchsian. On the other
hand, we have shown, see e.g.~\cite{gks}, that under very general
circumstances, the equation itself can be reduced to a nonlinear
Fuchsian ODE or PDE, without linearizing. We have also shown that the
initial-value problem for Fuchsian PDE can be solved in both the
analytic and non-analytic cases. The Chazy equation itself can be
reduced to Fuchsian form in several different ways, see the proofs of
theorems 7 and 8. The Fuchsian algorithm is not restricted to
`integrable' problems, but is in fact useful in analyzing singularity
formation in more general nonlinear wave equations, see
\cite{inversibilite,syd,gowdy}.  Of course, in the linear case, none
of the above issues arises: if a linear equation has solutions with
leading order $x^{\nu_1}$, $x^{\nu_2}$,\dots, the general solution is
simply a linear combination of these, and its singular behavior is
apparent. Confluence of singularities of linear Fuchsian equations
involves varying the coefficients of the equation, rather than
perturbing one solution of a fixed equation.

\subsection{The method of pole expansions}

The principle of the method is familiar: given an equation,
substitute for the unknown $u$ a series of the form:
\begin{equation}\label{1}
u = x^\nu\sum_{j\geq 0} u_j x^j,
\end{equation}
and identify like powers of $x$. This determines first $\nu$ and $u_0$,
hence the {\em leading balance} $u\sim u_0 x^\nu$. The other coefficients
are usually determined by a recurrence relation of the form
\begin{equation}
Q(j) u_j = F_j(u_0,\dots,u_{j-1}).
\end{equation}
The zeros of $Q$ are called {\em resonances}.

When $Q(j)=0$ and $F_j$ also vanishes, the coefficient $u_j$ is
arbitrary.  If $Q(j)=0$ but $F_j$ does not vanish, the series
(\ref{1}) must be corrected by the addition of logarithmic terms, and
one can predict their form rather precisely \cite{gks}. The
requirement that no logarithmic terms are needed sets rather strong
constraints on the equation; this observation was one of the
ingredients of the `Painlev\'e test' in its original form, because
many symmetry reductions of integrable equations do have this
property.  We do not discuss the modern status of this test, referring
the reader to recent reviews \cite{a-c,k-c,nlw,k-j-h} and their
references. The method can be extended to partial differential
equations, in which case it is known as the WTC method \cite{wtc,gks}.

Therefore, as far as this paper is concerned, the upshot is that
singularities of the form (\ref{1}) are expected to be stable under
perturbations, in the sense defined below, provided that the series
(\ref{1}) has the maximum number of arbitrary constants in it: any
nearby solution must have a singularity of the same type, with a
possible shift in location.

\subsection{Stability of singular behavior}

The stability of a singular solution will be defined by analogy with
the case of orbital stability of solitary waves in
translation-invariant problems: a solitary wave $u$ is (orbitally)
stable if any perturbation of the initial condition for a solitary
wave generates a solution which remains close to the orbit of $u$
under translations. Thus, in the case of the Korteweg-de Vries (KdV)
equation, an initial condition close to a one-soliton leads to a
solution which remains close to the set of all translates of this
soliton (see Strauss \cite{strauss,s-review}, Bona, Souganidis and
Strauss \cite{b-s-s} for the KdV case, and their references; further
results for KdV-like equations are also found in \cite{p-w}).

We encounter a comparable situation for movable singularities of
differential equations: if an equation is translation-invariant, we
expect that a small perturbation of any solution, in a domain away from the
singularity, will in general result in a {\em shift in the singularity
location}. This should be distinguished from a possible change in the
type of singularity.

As an example with a simple closed-form solution, consider the equation
\[
du/dx=2xu^2,
\]
and let us focus on real solutions to fix ideas; a similar discussion
could be made for complex solutions as well.
The general solution is
\[
u(x)={1\over c-x^2},
\]
where $c$ is a real constant, and consider small values of $c$. If $c$
is positive, we have two stable single poles: a slight change in the
value of, say, $u(1)=1/(c-1)$, results in a small change in the value
of $c$, hence a small shift of the pole.  However, if $c=0$, we
obtain an unstable double pole for $x=0$, because a small change in
the value of $u(1)$ causes the pole either to disappear, or to break
up into two simple poles, depending on whether $c$ becomes negative or
positive.

In the Chazy case, we will be interested in the stability of exact
solutions of the form $a/x$ in some fixed disk $D$ around the origin.
Fix also some nonzero value $\xi$ in $D$.  We will say that the
singularity is stable if {\em any} solution with Cauchy data at $\xi$
close to those of $a/x$, must have an expansion
\[
a/(x-x_0)+\sum_{j\geq 0} u_j (x-x_0)^j
\]
valid in $D$. We could allow logarithmic terms in the series, but they
will not be needed. We therefore require a simple pole to perturb to
another simple pole. An example of an unstable situation would be the
case in which a small change in the data at $\xi$ causes the pole to
turn into a natural boundary, or to split into several poles.

\subsection{The Chazy equation}

We recall here some background information on eq.~(\ref{chazy}).
The Chazy equation came up in the course of Chazy's
extension of Painlev\'e's program to third-order equations \cite{c1,c2,c3}.
It is actually one of the `class XII' equations (\ref{xii}).
It is closely related to a system considered by Halphen,
and its general solution can be parametrized using the solutions of a
hypergeometric equation if $k>0$, and the Airy equation if $k=0$ (see
Clarkson and Olver \cite{c-o}, Ablowitz and Clarkson \cite{a-c},
and their references).

The modern interest in this equation comes from the fact that it
arises as a reduction of self-dual Yang-Mills equations with an
infinite-dimensional gauge group \cite{a-c-c,a-c,t}. It arises in
connection with one of the special reductions of Einstein's equations
in a Bianchi IX space-time.  It can be given a commutator
representation. A particular solution is
\[
{1\over 2} {d\over dx}\ln\Delta(x),
\]
where $\Delta(x)$ is the discriminant modular form (see Takhtajan
\cite{t}, Bureau \cite{b1,b2}, and Chazy \cite{c3} for the relation to
modular forms, and Koblitz \cite{koblitz} for background material on
modular forms). This solution has the real axis as a natural boundary.
It generates a three-parameter family of solutions through
an SL(2) action which maps solutions to solutions: more precisely, if
$y(x)$ is a solution, so is
\begin{equation}\label{sl2}
{ad-bc\over(cx+d)^2}y({ax+b\over cx+d}) - {6c\over cx+d}
\end{equation}
for any choice of the complex parameters $a$, $b$, $c$ and $d$,
subject to $ad-bc\neq 0$. There are effectively only three parameters,
since scaling the parameters by a common factor does not generate a
new solution. The transformed solution has, in general, a circular
natural boundary. The relation of the Chazy equation to the
Schwarzian derivative and invariants of SL(2) actions was elucidated
by Clarkson and Olver \cite{c-o}, who showed how to use the group
action to obtain the general solution, despite the fact that the
symmetry group is not solvable.

On the other hand, the method of pole expansions generates exact
solutions which have no natural boundary at all, namely
\[
-{6\over x-x_0} + {A\over (x-x_0)^2},
\]
where $A$ is arbitrary. This is a two-parameter deformation of the
solution $-6/x$.

The problem is to decide the stability of the solution $-6/x$.

A similar problem arises for the general class XII equation, for which we
have the solutions \cite{c3,s-l}:
\[
{k-6\over 2(x-a)} - {k+6\over 2(x-b)},
\]
where $a$ and $b$ are arbitrary. One recovers the solution $-6/x$ by
confluence of $a$ and $b$. The resonances corresponding to the
singularities $a$ and $b$ are $(-1,1,k)$ and $(-k,-1,1)$ respectively.

\subsection{Earlier approaches}

{\bf 1. Linearization.}
The method of Fordy and Pickering \cite{f-p}
considers perturbations of the solution $-6/x$ of the form
\begin{equation}\label{pert}
-{6\over x} + \ep u_1(x) + \ep^2 u_2(x) +\dots
\end{equation}
and shows that one can compute the $u_k$ recursively by solving linear
equations involving the linearization of the Chazy equation at the
reference solution $-6/x$. This linearization reads
\[
(D+2)(D+3)(D+4)v=0,
\]
where $D=x d/dx$, and the corrections are sums of increasingly high
powers of $1/x$ and $x$.

The advantage of this method is that the new series does contain three
arbitrary parameters, which can be identified with the three
parameters required to describe the general solution of the linearized
equation.

The solutions of the linearization are more singular than $-6/x$, but
this does not mean that the equation itself has more singular
solutions: for example, we have
\[
-{6\over x-\ep} = -{6\over x}(1+\ep/x +\ep^2/x^2+\dots).
\]
Since the series converges only for $|x|>\ep$, the series does not
provide information on the behavior of this solution as $x$ approaches
zero. In fact, $x=0$ is a regular point for $\ep\neq 0$. Similarly, the
series (\ref{pert}) is not expected to converge for $x$ and $\ep$
small without any further restriction, since this would predict that
pole singularities perturb to Laurent expansions defined in a full
neighborhood of $x=0$, and such expansions do not allow for the
formation of a small natural boundary near the origin.

{\bf 2. Exponentially small corrections.}
Kruskal considered the general solution as a perturbation of the solution
$-6/x+A/x^2$. In this case, the linearized equation reads
\[
(x(D+4)-2A)(D+2)(D+3)v=0,
\]
where again $D=x d/dx$, and has therefore an {\em exponential solution}
which involves $\exp(-2A/x)$. The appearance of non-Fuchsian terms
was to be expected from the fact that the leading balance, for a
second-order pole, does not involve the top-order derivatives. He
derived systematically a representation of the general solution in the
form
\[
 -{6\over x} + {A\over x^2}(1+k\exp(-2A/x)+\dots),
\]
which, after including a parameter for translations, contains three
parameters, but is defined at best in a sector in which the exponential
is small. This sector has maximal angle $\pi$, and this restriction
corresponds to the fact that the solution cannot be continued around the
singularity at $x=0$ in general, but has a natural boundary which is a
circle (or a line) through $x=0$.

This representation can be checked directly in this special case using
the general solution \cite{a-c}. One can see by inspection
that as $A$ approaches zero, the natural boundary is a small circle
which shrinks to a point.

However, it is not clear how to account for resonance $-3$ in this
context, even though the solution thus obtained does contain three
arbitrary parameters.  Indeed, if we differentiate the expansion with
respect to $A$ or $k$ and set $A=0$, we cannot generate a function
with a fourth-order pole at the origin, as was possible in the
linearization approach.  As we will see, it is possible to recover the
fourth-order pole by using a {\em simultaneous variation} of all three
parameters.

{\bf 3. Remarks.}

Two other approaches to the interpretation of missing parameters in
pole expansions are as follows.

(1) The fact that some equations have two Painlev\'e series with
leading orders of the form $a/x$ and $na/x$ where $n$ is an integer
suggests that the second series results from the confluence of $n$
singularities of the first type (see Adler and van Moerbeke
\cite{a-m}, Ercolani and Siggia \cite{e-s}, Flaschka {\em et al.,}
\cite{f-n-t}, among others). However, as noted above, it remained
difficult to obtain an analytical representation of confluence,
because the pole expansion near one of the singularities has a
shrinking radius of convergence.

(2) It is possible to seek solutions in powers of $s=1/x$ rather than
$x$. In that case, negative resonances become positive resonances of
the equation in the $s$ variable. Such a series already appears in
Chazy \cite{c3}. However, this solution is only defined for $x$ large,
and there may be one or more natural boundaries which separate the
domain of existence of this solution from the origin. Also, this
interpretation is not applicable if there are both positive and
negative resonances.

\subsection{Results of this paper}

The necessary reconciliation of the linearization and exponential
approaches, for equations (\ref{chazy}) and (\ref{xii}) for $k=2$, 3,
4, 5, is obtained through the results below.

{\bf a.} (Theorems 1 and 9) Pole expansions with the maximum number of
arbitrary parameters represent stable singularities. This applies to
the solution $(k-6)/2x$.

{\bf b.} (Theorems 2--4) Any equation with the group invariance
property of the equation (\ref{sl2}) has a
one-parameter family of solutions $y(x;\ep)$ such that
\[
y(x;0)=-{6\over x};\qquad {\pa y\over \pa\ep}(x;0)=
{\mbox{const.}\; x^{-1+r}}
\]
for each negative resonance $r$ (i.e., for $r=-1$, $-2$ or $-3$).  The
perturbative expansion follows. This argument applies to
both (\ref{chazy}) and (\ref{xii}).

{\bf c.} (Theorem 5) Conversely, if an equation has the solution
${-6/x}$, it cannot have the SL(2) action (\ref{sl2}) if the
resonances do not include $-1$, $-2$ and $-3$. The existence of a full
expansion of $y(x;\ep)$ also follows from our proof.

{\bf d.} (Theorem 6) If there is a solution with branching involving
$x^k$, the group invariance proves the existence of pole-like
singularities with resonances $-1$ and $-k$.

{\bf e.} (Theorem 7) In the case of (\ref{chazy}), as $\ep$ increases
from zero, the isolated pole is unstable and goes into a circular
natural boundary of small radius. However, the general solution is
still described in terms of a convergent series of exponentials; this
follows from the fact that even though the linearization of the
equation can be non-Fuchsian, there still is a reduction to a
nonlinear Fuchsian equation.

{\bf f.} (Theorem 8) In the case of (\ref{xii}), for $k=2$, 3, 4 and 5, the
isolated pole does not turn into a natural boundary, but rather splits
into a finite number of poles. However, the confluence pattern is
restricted: all poles except one at most must coalesce. If all
confluence patterns had been allowed, the resonances of the solution
$-(k+6)/2x$ would have included all the negative integers from $-k$ to
$-1$. The confluence is described analytically by representing the
solution in terms of $u'/u$, where $u$ is analytic in a fixed domain
which includes both coalescing singularities.

\smallskip

This paper was motivated by M. D. Kruskal's questions on the relation
between the nonlinear Fuchsian approach to WTC expansion in
\cite{gks,nlw}, and the problem of negative resonances.  I am also
grateful to him for sharing with me his own results on this problem.


\section{Nonlinear Fuchsian equations}

We collect here a few results on Fuchsian equations which are used
later.  We also include a proof of the stability of polar
singularities with the maximum number of arbitrary coefficients in
their expansion (\ref{1}), which is similar in spirit to, but
considerably simpler than the result of \cite{inversibilite}, because
of the fact that we are dealing here with an ODE rather than a PDE.

\subsection{Existence results for Fuchsian ODE}

A nonlinear Fuchsian equation has the general form
\begin{equation}\label{fuchs}
P(D)u(x)=xF[x,u,Du,\dots,D^{m-1}u],
\end{equation}
where $D=xd/dx$, and $P$ is a polynomial of degree $m$.  There are
similar definitions and results for systems, but they will not be
needed here.

If the zeros of $P$ all have negative real parts, there is precisely
one solution of this equation which vanishes for $x=0$, and it is
given by a convergent series in powers of $x$ near $x=0$. The proof
is by iteration in a space of analytic functions, and is in this sense
constructive. In particular, it could in principle be used to
generate the coefficients of the expansion of the solution, but it is
in practice quite cumbersome to do so.

The convergence of WTC expansions follows from the fact (see
\cite{nlw}) that it is possible to allow $u$ to depend on additional
`transverse' variables, provided that spatial derivatives terms are
multiplied by appropriate powers of $t$, see \cite{syd,gks}.  This
condition is satisfied in a remarkably large number of cases.

\smallskip

We will meet in section 3 an equation of the form
\begin{equation}\label{or}
P(D)Du=xF,
\end{equation}
where $P$ and $F$ are as above. To reduce this situation to the the case
of (\ref{fuchs}), let us write $u=a+xv(x)$, where $a$ is arbitrary.
One can write
\[
xF=xF[0,a,0,\dots] + x^2G[x,v,Dv,\dots,D^{m-1}v],
\]
for some function $G$.
The equation now becomes, after division by $x$,
\[
Q(D)v:=P(D+1)(D+1)v=F[0,a,0,\dots]+xG,
\]
where the zeros of $Q$ all have negative real parts. We now replace
$v$ by $v-b$, for a suitable constant $b$, to annihilate the first
term on the right-hand side. In this way, we reduced the problem to an
equation of the form (\ref{fuchs}), and we conclude that there is a
unique solution $v$ which vanishes at the origin. Therefore, for any
$a$, there is a unique analytic solution of (\ref{or}) which satisfies
$u(0)=a$.

\subsection{Stability and parameter dependence}

We now prove a stability result for pole singularities of an equation with
the maximum number of coefficients in their pole expansions. This result
makes rigorous the intuitive argument to the effect that a series which
contains as many free parameters as there are Cauchy data must represent
the general solution locally.

To prove the result, one must show that these parameters are not
redundant. We achieve this by a reduction to the implicit function
theorem. An example of a redundant parametrization is the two-parameter
family of series:
\begin{equation}\label{ex}
u(x;\ep,\eta)=\sum_{j\geq 0} {\eta^j\over (x-\ep)^{j+1}}.
\end{equation}
The parameters $\ep$ and $\eta$ are
redundant, because $u(x;\ep,\eta)=1/(x-\ep-\eta)$: the pairs
$(\ep,\eta)$ with the same value of $\ep+\eta$ all define the same function.

To fix ideas, let us consider an autonomous equation of the form
\begin{equation}\label{eq-f}
(d/dx)^mu=f(u,\dots,u^{(m-1)}),
\end{equation}
where $f$ is, say, a polynomial.

Let $u(x-x_0,c_1,\dots,c_{m-1})$ be a family of solutions, which depends
analytically on $(x_0,c_1,\dots,c_{m-1})$ for $|x_0|$ and $|c_k|<a$ and
$0<|x-x_0|<b$, for some positive $a$ and $b$.
We have the following result:
\begin{th}
Assume that
$\pa u/\pa x_0$, $\pa u/\pa c_1$,\dots, $\pa u/\pa c_{m-1}$
form a linearly independent set of solutions of the linearization of
(\ref{eq-f}). Then $u(x-x_0,c_1,\dots,c_{m-1})$ is a local representation
of the general solution. In particular, the assumption holds for any pole
expansion with the maximum number of parameters if $\nu\neq 0$.
\end{th}

{\em Remarks:}
(1) If $\nu=0$, we are in the case of the Cauchy problem, and the series
for the solution contains $m+1$ parameters, namely the location of the
initial point and the $m$ Cauchy data. These data are clearly redundant.

(2) In the case of (\ref{ex}), the representation is redundant because
$\pa u/\pa \ep=\pa u/\pa \eta$.
\smallskip

{\em Proof:} Consider the reference solution $U=u(x,0,\dots,0)$ for
definiteness. Given any point $x_1$ with $0<|x_1|<b$, we consider
\[
\varphi : (x_0,c_1,\dots,c_{m-1}) \mapsto
          (u(x_1), u'(x_1), \dots, u^{(m-1)}(x_1)),
\]
where $u(x_1)=u(x_1-x_0,c_1,\dots,c_{m-1})$, and similarly for the
derivatives of $u$. Applying the inverse function theorem to this map
near $(x_0,0,\dots,0)$, we conclude that any set of Cauchy data close
to the data of $U$ at $x_1$ coincides with the Cauchy data of a member
of our family, QED.

To prove that the linear independence condition holds in the context
of the theorem, it suffices to consider the family
$u(x-x_0,c_1,\dots,c_{m-1})$, where the $c_l$'s are the arbitrary
coefficients in the expansion of $u$. The functions $\pa u/\pa x_0$,
$\pa u/\pa c_l$ are derivatives of families of solutions, and are
therefore themselves solutions of the linearized equation. It is easy
to see that these derivatives all have different leading behaviors at
$x=x_0$, and are therefore linearly independent, QED.


\section{The transformation formula and negative resonances}

\subsection{Statement of results}
Consider any equation which admits the transformation formula
(\ref{sl2}). Assume that any uniform limit of analytic solutions is
also a solution. This assumption is clear for ODEs; it will allow us
to extend (\ref{sl2}) to some cases when the transformation
$(ax+b)/(cx+d)$ is non-invertible, by viewing it as a limit of
invertible transformations.

Let $y(x)$ be any solution. We wish to prove that
there are families of solutions $y(x;\ep,r)$ such that
        \[
        y(x;0,r)=-6/x\mbox{ and } {dy\over d\ep}(x;0,r)=x^{-1+r}
        \]
for $r=-1$, $-2$ and $-3$. This will account for the three `negative
resonances.' It will be apparent from the proofs that $y(x;\ep;r)$ can in
fact be expanded to higher order, and that the coefficients of the
higher-order terms are increasingly more singular in $x$.

We find that the construction is possible provided that it is possible
to prescribe $y$, $y'$, $y''$ arbitrarily at one point. This
construction precisely fails for the non-generic solutions $-6/x +
A/x^2$. The fact that the resonance structure can be derived on the
sole basis of the representation formula implies conversely that if we
have an equation with a different resonance structure, it {\em cannot}
admit the transformation formula (\ref{sl2}).

Fix a solution $y(x)$ which is analytic near $x=0$. Consider the family
\begin{equation}
        y(x;\ep) = -{6\over x-\eta}
                    + {\mu\over (x-\eta)^2}y({-\mu\over x-\eta}),
        \label{mu}
\end{equation}
where $\eta$ and $\mu$ depend on $\ep$, and are assumed to be small as
$\ep\to 0$. This is a special case of the transformation (\ref{sl2}).

Our results are as follows.

\begin{th}
        If $\mu y(0)-6\eta = \ep$, and $\mu$ and $\eta$ are proportional to
        $\ep$,
        \[
        y(x;0)=-6/x \mbox{ and } {dy\over d\ep}(x;0)=1/x^2.
        \]
\end{th}

\begin{th}
        Assume $\mu y(0)-6\eta = 0$, but $6y'(0)-y(0)^2\neq 0$. Then, if $\mu$
        and $\eta$ are both proportional to $\ep^{1/2}$, we have
         \[
        y(x;0)=-6/x \mbox{ and } {dy\over d\ep}(x;0)= c/x^3,
        \]
        where $c\neq 0$.
\end{th}

\begin{th}
        Assume $\mu y(0)-6\eta = 0$ and $6y'(0)-y(0)^2 = 0$, but
        $y''-yy'+y^3/9\neq 0$ when $x=0$. Then, if $\mu$
        and $\eta$ are both proportional to $\ep^{1/3}$, we have
         \[
        y(x;0)=-6/x \mbox{ and } {dy\over d\ep}(x;0)= c/x^4,
        \]
        where $c\neq 0$.
\end{th}

\begin{th}
        If an equation of order three or higher, to which the Cauchy existence
        theorem applies, admits the special solution
        $y=-6/x$, and if the linearization of the equation at this
        solution does not
        have $1/x^2$, $1/x^3$ and $1/x^4$ among its solutions, then the given
        equation cannot admit the SL(2) action (\ref{sl2}).
\end{th}

The restriction that the solution $y(x)$
be analytic is essential. In fact, we can
obtain quite different results if $y$ admits branching:
\begin{th}
        If there is a solution of the form $y(x)=x^{-1}h(x^k)$, where $h$ is
        analytic, $k>0$, and $h(0)=(k-6)/2$, there exist two families of
        solutions, $y_1(x;\ep)$ and $y_2(x;\ep)$, such that
        \[
        y_1(x;0)=y_2(x;0)=-{k+6\over 2x},
        \]
        and
        \[
        {dy_1\over d\ep}(x;0)= c/x^2 \mbox{ and }{dy_2\over d\ep}(x;0)= c/x^{k+1}.
        \]
\end{th}

\subsection{Remarks}

\hskip\parindent%
{\bf 1.}
If we take $y=-6/(x-x_0)$, theorems 2 and 3 do not apply.

{\bf 2.}
If we take $y=-6/(x-x_0)+A/(x-x_0)^2$, theorem 3 applies,
but theorem 4 does not.  Indeed, in this case,
\begin{equation}\label{2}
y'-y^2/6=-A^2(x-x_0)^{-4}/6,
\end{equation}
and
\begin{equation}\label{3}
y''-yy'+y^3/9=(d/dx-(2/3)y)(y'-y^2/6)=A^3(x-x_0)^{-6}/9.
\end{equation}
It is therefore not possible to make (\ref{2}) vanish without
having $A=0$---in which case (\ref{3}) vanishes as well.
One can rephrase the assumption in theorem 4 by saying that we
require $y'=y^2/6$ but $y''\neq y^3/18$, for $x=0$.

{\bf 3.}  Theorems 2, 3 and 4 all apply, for example, when there is a
solution for every choice of $y(0)$, $y'(0)$ and $y''(0)$. The result
therefore holds for any third-order autonomous equation, hence for
both equations (\ref{chazy}) and (\ref{xii}) which, as we show in
section 4, have completely different singularity structures.

{\bf 4.}
It follows from theorem 8 below that theorem 6 applies to
equations (\ref{xii}). A result similar to theorem 5 could of course be
stated for this situation.

{\bf 5.}
An example to illustrate theorem 5 is the equation
\[
y'''=2yy''-3y^{\prime 2}+cy'(6y'-y^2)
\]
which has the solution $-6/x$, but where the resonance equation is
$(r+1)[(r+2)(r+3)-36c]=0$. $-2$ and $-3$ are therefore both resonances
only if $c=0$. We conclude, without computing the symmetry group of
the equation, that this equation does not admit the transformation law
(\ref{sl2}) if $c\neq 0$.

\subsection{Proofs}

Let us begin with a computation which is used in the proofs of the first
three theorems. Any solution $y(x)$ generates the one-parameter family of
solutions
\[
y(x;\ep) = -{6\over x-\eta(\ep)}
+ {\mu(\ep)\over (x-\eta(\ep))^2}y(-{\mu(\ep)\over x-\eta(\ep)}).
\]
If $x$ is fixed and nonzero, and if $\mu$ and $\eta$ are small as $\ep\to
0$, we can expand this solution in the form
\[
\begin{array}{rcl}
y(x;\ep) & = & \displaystyle{
                  {-6\over x} + {\mu y-6\eta\over x^2}
                 +{\mu(2\eta y-\mu y')-6\eta^2\over x^3} }\\
         &   &
 \mbox{}+x^{-4}[-6\eta^3+\mu(3\eta^2 y-3\eta\mu y'+{\mu^2\over 2}y'')]\\
         &   &\mbox{} +O(\eta^4, \eta^3\mu, \eta^2\mu^2, \eta\mu^3,\mu^4),
\end{array}
\]
where $y$, $y'$,\dots\ stand for $y(0)$, $y'(0)$,\dots

Any such family has the property that $y(x;0)=-6/x$. Furthermore, it is
clear that the above expansion could be pushed to all orders, and that
the coefficients of the higher order terms contain higher and higher
powers of $1/x$.

{\bf Proof of Th.~2:}
If we take $\mu$ and $\eta$ proportional to $\ep$, in such a way that
$\mu y(0)-6\eta\sim\ep$, we have $\pa y/\pa\ep=1/x^2$ for $\ep=0$.

{\bf Proof of Th.~3:}
If we take $\mu$ and $\eta$ proportional to $\ep^{1/2}$, in such a way that
$\mu y(0)-6\eta=0$, and if $y$ is such that $6y'(0)\neq y(0)^2$,
we have $\pa y/\pa\ep=\mbox{const.}/x^3$ for $\ep=0$.

{\bf Proof of Th.~4:}
If we take $\mu$ and $\eta$ proportional to $\ep$, in such a way that
$\mu y(0)-6\eta=0$, and assume that $6y'(0)-y(0)^2=0$, but
$y''-yy'+y^3/9\neq 0$ for $x=0$, we find that $\pa
y/\pa\ep=\mbox{const.}/x^4$ for $\ep=0$.

This proves theorems 2, 3 and 4.

{\bf Proof of Th.~5:}
Consider an equation $F[u]=0$ of order three or higher with such a group
action. Solving the
Cauchy problem, we can construct solutions to which each of theorems 1, 2
and 3 apply. Consequently, there are differentiable families of solutions
$y(x;\ep)$ as in these theorems. Since $F[y(x;\ep)]$ is identically zero,
we have
\[
0={d\over d\ep}F[y(x;\ep)]{\vrule height 10pt depth 10pt width .4pt
}_{\;\ep=0}
=F'[-6/x]({dy\over d\ep}(x;0)),
\]
where $F'$ denotes the linearization of $F$. We conclude that this
linearized equation must admit the three solutions $1/x^m$, $m=2$, 3
and 4.  If these three functions do not solve the linearization, there
can be no such group action, QED.

The specific coefficients of the group action are not essential
to the result: only the existence of an expansion of families of
solutions matters.

{\bf Proof of Th.~6:}
The solution $y(x)$ in the statement of the theorem is constructed in
theorem 8.

From $y$, we construct the one-parameter family:
\[
y_2(x;\ep)=-{6\over x}-{1\over x}h({\ep\over x^k}),
\]
using (\ref{sl2}) for the inversion $x\mapsto \ep^{1/k}/x$.

Letting $\ep\to 0$, we find that $-(6+h(0))/x=-(k+6)/2x$ must be a solution.
We now define
\[
y_1(x;\ep)=-{k+6\over x-\ep}.
\]
The properties listed in the theorem are now readily verified.


\section{Instability of isolated poles}

\subsection{Results}

Even though the construction of the perturbation expansion of
solutions close to $-6/x$ can be made solely on the basis of the group
action on solutions, the singularities which arise by perturbation of
simple poles are different for (\ref{chazy}) and (\ref{xii}). We know
that perturbation series near a single pole do not allow an analytical
description of confluence phenomena. However, even though a
function such as $(x-a)^{-1}+(x-b)^{-1}$ is not jointly analytic in
$x$, $a$ and $b$ small, it is the logarithmic derivative of
$(x-a)(x-b)$ which is perfectly well-behaved. More generally, we show
that a Cole-Hopf transformation provides an analytical description of
confluence phenomena in the Chazy equation.

More precisely,

\begin{th}
        For any constant $a$,
        equation (\ref{chazy}) has precisely one solution of the form
        $y(x)=u'/2u$ with
        \[
        u(x) = e^x(1+e^xw(e^x)),
        \]
        where $w$ is analytic when its argument is small, and
        $w(0)=a$.  Using transformations (\ref{sl2}), this solution
        generates a one-parameter family of perturbations of $-6/x$,
        with a natural boundary shrinking to a point as the parameter
        vanishes. Their asymptotics at the boundary are those
        suggested by the method of exponential corrections.
\end{th}

For equation (\ref{xii}), we have

\begin{th}
    Let $a$ be a constant.
        For $k\neq 0$ or $1$, equation (\ref{xii}) has a unique solution
        of the form
        \[
        y(x)=x^{-1}h(x^k),
        \]
        where $h$ is analytic when its argument is small,
        $h(0)=(k-6)/2$, and $h'(0)=a$. If $k=2$, $3$, $4$ or $5$, this
        solution is rational. Using transformations (\ref{sl2}), this
        solution generates a one-parameter family of perturbations of
        $-6/x$, where all poles, except possibly one, cluster at the
        origin as the parameter vanishes.
\end{th}

Thus, $-6/x$ is unstable; the next result shows that $(k-6)/2x$
is stable, but $-(k+6)/2x$ is unstable:
\begin{th}
  For $k=2$, $3$, $4$, $5$, there is a three-parameter family of
  solutions of (\ref{xii}) which contains the solution $(k-6)/2x$.
  These parameters are in correspondence with the Cauchy data at a
  nearby regular point. Solutions with leading term $-(k+6)/2x$ on the
  other hand are unstable under perturbation: they arise from the
  confluence of all singularities save one.
\end{th}

\subsection{Remarks}

{\bf 1.}  Since equation (\ref{ch-infty}) below admits the
discriminant modular form $\Delta$ as a special solution \cite{r},
Theorem 7 provides a proof that $\Delta$ is entirely determined by the
first two terms of its expansion in powers of $q=\exp(2\pi i x)$.  Its
reduction to Fuchsian form implies in particular a (perhaps new)
recurrence relation on the coefficients of $\Delta$, that is, on the
Ramanujan $\tau$ function.  On the other hand. our proof does not make
use of any properties of modular forms, and therefore suggests that
some of the phenomena found in the Chazy equation are of wider
significance.

{\bf 2.}
The existence of rational solutions and the observation that a
Cole-Hopf transformation simplifies some of the issues is found in Chazy
\cite{c3}.

\subsection{Proofs}

For clarity, some tedious but straightforward computations have been
omitted; we sometimes found it convenient to perform some of the
verifications using a symbolic manipulation package.

{\bf Proof of Th.~7:}
Let $y$ be a solution of (\ref{xii}).
Let
\[
y={u'\over 2u}.
\]
We find that $u$ satisfies:
\begin{equation}
   \label{ch-infty}
   u^3u^{(4)}-5u^2u'u^{(3)}-{3\over 2}u^2u^{\prime\prime 2}
   + 12 uu^{\prime 2}u''-{13\over 2}u^{\prime 4}=0.
   \end{equation}
If $u$ is a solution, so is
\[
(cx+d)^{-12}u({ax+b\over cx+d}).
\]
Since $e^x$ is an exact solution of this equation, we seek solutions of
the form $e^xv(e^x)$. Note that $u=\exp(2bx)$ leads to $y=b$, that is,
to constant solutions of (\ref{chazy}).

We make change of variables $z=e^x$, and let $D:=zd/dz=d/dx$.
This turns (\ref{ch-infty}) into a Fuchsian equation for $v(z)=u(z)/z$:
\begin{eqnarray*}
\lefteqn{v^3(D+1)^4v-5v^2(D+1)v(D+1)^3v-{3\over 2}v^2(D+1)^2v}\\
  &\mbox{} + 12 v[(D+1)v]^{2}(D+1)^2v-{13\over 2}[(D+1)v]^{4}=0.
\end{eqnarray*}
Letting $v(z)=1+zw(z)$, we find that $w$ satisfies an equation of the form
\[
(D+1)^3Dw=zG[z,w,Dw,D^2w,D^3w].
\]
It follows that there is exactly one solution with $w(0)=a$, and that it
is given by a convergent series in $z$ near $z=0$.

Coming back to $x$, we have obtained a solution of the desired form,
given by a series of exponentials which converges at least
for Re$\;x < -\rho$ for some finite $\rho$. We know in fact that $\rho$
is equal to zero for the case $u=\Delta(ix)$. This completes the proof of
the claims regarding the family of exponential solutions.

Next, let us take the case when the solution $y$ has the real axis for a
natural boundary, to fix ideas.

Consider the transformation (\ref{sl2}) generated by
$x\mapsto \ep x/(x-i\ep)$, which maps the real axis to the circle
$(\Gamma_\ep)$ of center
$\ep/2$ and radius $\ep$. The solution $y(x)$ generates the one-parameter
family of solutions:
\[
y(x;\ep) = -{6\over x-i\ep}-{i\ep^2\over(x-i\ep)^2}
                               y({\ep x\over x-i\ep}),
\]
which are defined {\em outside} $(\Gamma_\ep)$. As $\ep\to 0$, we see
that the natural boundary shrinks to a point, and that the solutions
$y(x;\ep)$ converge, uniformly on any disk at positive distance from the
origin, to the solution $-6/x$.

However, the limits $\ep\to 0$ and $x\to 0$ do not commute; in fact,
$y(x;\ep)$ is not defined in a full neighborhood of $x=0$ for all small
values of $\ep$.

This completes the proof of theorem 7.

{\bf Proof of Th.~8:}
Let $y$ be a solution of (\ref{xii}).
Let
\[
y={k-6\over 2}{u'\over u}.
\]
We find that $u$ satisfies:
\begin{equation}
        uu^{(4)}-(k-2)u'u^{\prime\prime\prime}
                +{3k(k-2)\over 2(k+6)}u^{\prime\prime 2}=0.
        \label{k-eq}
\end{equation}
If $u$ is a solution, so is
\[
(cx+d)^{12/(6-k)}u({ax+b\over cx+d}).
\]
The first part of the theorem follows from general results on
nonlinear Fuchsian equations.  Let us seek $y$ in the form
\[
y(x)=x^{-1}(a+bz+w(z)z^2),
\]
where $z=x^k$, $a=(k-6)/2$ and $b$ is arbitrary. Letting $zd/dz=D$, we
find, after substitution into the equation and some algebra,
that $w$ satisfies an equation of the form
\[
(D+1)(k(D+2)+1)(k(D+2)-1)w=zF[z,w,Dw,D^2w].
\]
It follows that there is precisely one solution of the form $y=h(x^k)/x$
if we specify $h(0)=(k-6)/2$ and $h'(0)=b$. This proves the first part of the
theorem. If $b=0$, we find $w\equiv 0$.

Let us now focus on $k=2$, 3, 4, 5. In each case, there is a polynomial
solution of (\ref{k-eq}), which generates the desired solutions using the
SL(2) action \cite{c3}. In fact, we have
\[
u=(x-a_1)\dots(x-a_N),
\]
and
\[
y(x)={1\over 2}(k-6)\sum_{j=1}^N{1\over x-a_j}
    ={(k-6)\over 2x}\sum_{n\geq 0} {\sum_j a_j^n\over x^n},
\]
with $N=1+(k+6)/(6-k)=12/(6-k)$.
Note that $u$ is analytic near $x=0$ even when the $a_j$ tend to zero.
The relation between linearized solutions and possible confluence
patterns is given by the following:
\begin{lemma}
  If we can choose the pole locations such that $a_j=\ep^{1/m} b_j$,
  where $\sum_j b_j^q$ vanishes for $q<m$, but is nonzero if $q=m$,
  then
        \[
        y(x;\ep)=-{6\over x}+{\mbox{const.}\;\ep\over x^{1+m}}(1+o(1)).
        \]
\end{lemma}
In other words, we have a resonance at $-m$. Since the possible pole
locations are obtained by applying homographic transformations to the
zeros of a fixed function, not all pole configurations are possible.

The lemma follows by direct computation.

\smallskip

If all the poles are equal to zero, we recover $y=-6/x$; if they are
all zero except for one which we let tend to infinity, we obtain the
solution $-(k+6)/x$. If all poles but one are sent to infinity, we
obtain the solution $(k-6)/x$. It is apparent that the first two
solutions are unstable.

\smallskip

Let us now show that there cannot be any other type of confluence.
Assume that there is a family of homographic transformations,
depending on a parameter $\ep$, under which two distinct poles $a_1$
and $a_2$ tend to zero while two other poles $a_3$ and $a_4$ remain
fixed at nonzero (distinct) locations. We do not restrict the location
of any additional poles. The anharmonic ratio of $(a_1,a_2,a_3,a_4)$
tends to 1. But it is also independent of $\ep$; it is therefore
identically equal to 1. This implies that $a_1=a_2$ for all $\ep$: a
contradiction.  Therefore, if there is such a confluence, all poles
except one at most, must cluster at the same point.

{\bf Proof of Th.~9:} The stability statements have already been
proved in the course of the proof of the previous theorem. Since the
solution $(k-6)/2x$ has two positive resonances, namely 1 and $k$, we
expect to be able to conclude using theorem 1. There are in fact no
logarithms in the pole expansion, but this does not follow from
theorem 7, which only generates a one-parameter solution corresponding
to the resonance $k$: to generate the complete solution, we need to
check that the resonance 1 is compatible. It is convenient to do so
using the group action. More precisely, the solution $y=x^{-1}h(x^k)$
generates the solutions
\[
-{6\ep\over 1+\ep x}+{h(x^k/(1+\ep x)^k)\over x(1+\ep x)},
\]
which contain the additional parameter $\ep$. Adding the translation
parameter, we obtain a three-parameter family to which theorem 1
applies; this completes the proof.


\end{document}